\documentclass{aa}\usepackage{psfig}

\begin{document}


\title{Mid Infrared Polarisation of Ultraluminous
Infrared Galaxies \thanks{Based on observations with ISO, an ESA
project with instruments funded by ESA Member States (especially the
PI countries: France, Germany, the Netherlands and the United Kingdom)
with the participation of ISAS and NASA.}}

\author {R.~Siebenmorgen\inst{1} and  A.~Efstathiou\inst{2}}

\institute{
\inst{1} European Southern Observatory, Karl-Schwarzschildstr. 2, 
D-85748 Garching b.M\"unchen, Germany \\
\inst{2} Astrophysics Group, Blackett Laboratory, Imperial College of
Science Technology \& Medicine, Prince Consort Rd., London.SW7 2BZ }

\offprints{Ralf Siebenmorgen, \email{rsiebenm@eso.org}}

\date{Received May 25, 2001 / Accepted July XX, 2001}

\abstract{The mid infrared polarisation properties of four
Ultraluminous Infrared Galaxies (ULIRGs) have been investigated by
broad band filter observations with the ISOCAM instrument on board 
the Infrared Space Observatory (ISO). The wavelength region from 5 to
18\,$\mu$m was selected where the emission from the putative torus
peaks. We report detection of polarisation for all ULIRGs studied. The
fractional polarisation ranges from $\sim 3\%$ up to 8\%.  The highest
polarisation is recorded in Mrk~231 which has a clear AGN signature,
whereas the lowest is for Arp~220, which is generally thought to be
powered predominantly by star formation. We discuss the various
mechanisms that could give rise to the polarisation and conclude that
the most likely interpretation is that it is due to magnetically
aligned elongated dust grains. This is the same mechanism believed to
be operating in a number of galactic sources.  The position angle of
polarisation could give the projected magnetic field direction and
therefore constrain models for the formation of the tori.
}

\maketitle

\section{Introduction}


One of the most surprising discoveries of the Infrared Astronomical
Satellite (IRAS) was that of a local population of Ultraluminous
Infrared galaxies (ULIRGs) with far IR luminosities ($L_{\rm FIR} >
10^{12} L_\odot$) and space densities higher than those of quasars
(Sanders et al. 1988). The recent discovery of a large number of
submillimeter galaxies, which could be primeval ULIRGs, with SCUBA
(Hughes et al. 1998, Barger et al. 1998) and other far IR surveys
(Puget et al. 1999, Kawara et al.  1998, Efstathiou et al. 2000), has
renewed interest in this class of object.

 The origin of the extreme luminosity of ULIRGs, is still a matter of
controversy.  While it is generally agreed that the IR
luminosity is due to emission from dust that reprocesses optical and
UV light to longer wavelengths, it is still not clear whether the
primary source of radiation is a burst of star formation (e.g. Downes
\& Solomon 1998, Skinner et al. 1997) or dust-enshrouded accreting
black holes (Sanders et al. 1988, Soifer et al. 2000).

 A number of observational techniques have been employed in an attempt
 to settle this issue.  However, the high degree of obscuration makes
 traditional diagnostics such as optical and near IR (Goldader
 et al. 1995) emission lines difficult to interpret. Recently,
 mid IR spectroscopy which is less affected by obscuration,
 obtained with the ISO satellite, has shown that ULIRGs are
 predominantly powered by star formation (Genzel et al.  1998,
 Rigopoulou et al. 1999, Clavel et al. 2000). However, the optical
 depth of the dusty medium (believed to be in the form of a torus)
 that is obscuring the AGN may be optically thick to suppress even the
 mid IR lines.

Following the discovery of broad emission lines in the polarised flux
of the prototypical narrow-line active galactic nucleus (AGN) NGC1068
(Antonucci \& Miller 1985), a lot of effort has been directed towards
the study of the polarisation properties of galaxies (Heisler et al. 
1997). The generally accepted interpretation of this phenomenon is
that our view of the sub-parsec scale region of ionized gas producing
the broad emission lines is blocked by a geometrically and optically
thick dusty torus. In this situation the broad lines are detected
because they are reflected (and hence polarised) by free electrons
along the axis of the torus. The postulated dusty torus is the central
theme of the `unified model' for active galaxies which holds that a
lot of the diversity in their observed properties is the result of the
orientation of the observer with respect to the axis of the torus
(Antonucci 1993).

An important corollary of the obscuring torus hypothesis is that a
large fraction of the power emitted by an AGN is absorbed and
re-emitted by dust in the IR.  Models of the IR emission
from such tori have been presented by a number of authors (Pier \&
Krolik 1992, Granato \& Danese 1994, Efstathiou \& Rowan-Robinson
1995, Ruiz et al. 2001). The predicted emission peaks in the
$3-30 \mu$m range and shows spectral features at $9.7 \mu$m due to
silicate dust. However, since the torus emission is usually unresolved
it is difficult to separate the emission of the torus from that of
other dusty parts of the galaxy such as giant molecular clouds heated
by young stars (Kr\"ugel \& Siebenmorgen 1994).

Until large IR telescopes/interferometers for directly imaging the
torus become available its properties can only be probed by indirect
methods, a very powerful of which promises to be IR polarimetry
(Efstathiou et al.  1997). IR polarisation can of course be produced
by non-thermal processes (synchrotron), and we discuss this later, but
what is of more interest to us here is polarisation produced by the
dichroic absorption (or emission) by non-spherical magnetically
aligned grains. This effect is similar to the one that gives rise to
the polarisation in the optical by interstellar grains aligned by the
galactic magnetic field (Hall 1949, Hiltner 1949).
 
If dust grains are similarly aligned in the dusty torus by the
magnetic field, which in some models is thought to be a prerequisite
for its formation (Krolik \& Begelman 1988, Koningl \& Kartje 1994),
then we expect the radiation emitted (and self-absorbed) by the torus
to be polarised (Efstathiou et al. 1997). The region that produces the
IR polarisation, the torus, is therefore different from the region
that produces the optical and UV polarisation.  The effect of
polarisation due to magnetically aligned dust particles has been
observed so far in a few galactic environments such as the Galactic
Centre (Hildebrand 1989), molecular clouds (Novak et al.~1989, Clemens
et al.~1999), star forming regions (Gonatas et al.~1990, Leach et
al.~1991, Smith et al.~2000), protostars (Siebenmorgen \& Kr\"ugel
2000) and prestellar cores (Minchin et al.~1995, Ward-Thompson et
al.~2000). Extended polarisation was recently reported for two
external galaxies: M82 (Greaves et al.~2000) and NGC1808 (Siebenmorgen
\& Kr\"ugel 2001). The latter authors argue against mid IR
polarisation due to small grains such as PAHs or nanometre--sized
silicates.

With the expectation of detecting the IR polarisation from dusty tori
in ULIRGs, if they exist, we embarked on a program of mid IR
polarimetry with the ISOCAM instrument (Cesarksy et al. 1996) on the
Infrared Space Observatory (ISO, Kessler et al. 1996). During the ISO
mission four of the ULIRGs in the IRAS Bright Galaxy sample ($60 \mu$m
flux density $S(60)> 5$\,Jy; Sanders et al. 1988) could be
observed. Targets were selected mainly on visibility constraints so
that they represent a random subset of the Sanders sample.

\section{Observations}

ISOCAM (Cesarsky et al. 1996) polarimetric imaging of the ULIRGs
studied was set up according to the observing template CAM05
(Siebenmorgen et al.~1996).  We used a 2$\times$2 raster with a raster
step size of 36$''$ and a 6$''$ pixel field of view.  The read out
time of each exposure was 2.1s. We took initially 60 exposures on the
source to stabilize the detector and then performed a raster through
the free hole of the entrance wheel.  On each raster position of the
hole measurements 15 exposures were read. Then three raster were taken
by rotating the polarisers on each raster position. For each
combination of polariser and raster position 23 exposures were read.
The polariser rasters were repeated in 3 times.  This procedure gives
a total of twelve independent measurements for each polarisers.

We present broad band filter observations LW2, LW10 and LW3, centred
at 6.7, 12.0 and 14.3\,$\mu$m, at resolution of 5.4$''$, 9.6$''$ and
11.4$''$, respectively.  The data are reduced with the ISOCAM
interactive analysis system (CIA version 4.0, Ott et al. 1996).  Only
basic reduction steps, such as dark current subtraction, removal of
cosmic ray hits, transient correction and flat fielding are applied
(see Siebenmorgen et al.~(2000) for more details on ISOCAM
data). Photometry is done by integrating the flux in an aperture
radius of 6 pixels. A background estimate is derived from the mean
flux in a 4 pixel wide circular annulus 2 pixels away from the source
aperture. The same procedure is applied on a model point spread
function (Okumura et al. 1998) to correct the derived source fluxes
for losses in higher airy rings.  The dead column of the LW --
detector was masked out. Instrumental polarisation is corrected by
applying the polarisation weight factors as described by Siebenmorgen
(1999).

For each measurement the polarised signal is found to be consistent
between the cycles. The average of all cycles gives the final
Stokes vector. The (unpolarised) standard star HIC085371 was also
observed with the same strategy as a test of our observing and data
reduction procedure. We find a residual degree of polarisation of $p =
0.45 \pm 0.4 \%$ for this star, which reflects the limit of our
processing technique and calibration uncertainty. This observing mode
and data reduction procedure has also been used to detect a 20\%
polarisation at $14.3 \mu$m in the Crab nebula at a position angle
consistent with optical and radio measurements (Tuffs et al. 1999),
thus providing a further test of the methods applied. 

Results of the ISO observations together with near IR polarimetry by Jones
\& Klebe (1989) of the four ULIRGs are summarised in
Table~\ref{pol.tab}.  At the resolution of ISO the objects appear
point like. The corresponding observing sequence or TDT numbers read
for target: Mrk~231 (56800101, 56800201), Arp~220 (61900201, 61900301,
61900401), IRAS~15250+3609 (57400701, 57400801, 57400901) and Mrk~273
(52700101, 52700201, 52700301), respectively.

  \begin{table}[h!tb] 
\begin{center} 
\caption {Infrared polarisation of ULIRGs. Near IR data by Jones \&
Klebe (1989). ISO data this work.}
\label{pol.tab}

\begin{tabular}{|l|l|l|l|l|}  
\hline 

Object           &   Band   &   $F$        &  $p$   &  $\theta$  \\ 
	       & $\mu$m         &   mJy         &    \%             & deg \\
\hline
Mrk~231         &  J             &               &  0.77$\pm$0.14    &  102$\pm$6     \\
               &  H             &               &  0.62$\pm$0.12    &   95$\pm$6     \\
               &  K             &               &  0.46$\pm$0.08    &  107$\pm$5     \\
	       &  12.0          &   989$\pm$7   &  8.6$\pm$0.9      &  123$\pm$4     \\
               &  14.3          &  1719$\pm$10  &  6.7$\pm$0.9      &  126$\pm$4     \\
               &                &               &                   &                \\  
Arp~220         &  J             &               &  1.15$\pm$0.25    &  58$\pm$7     \\
               &  H             &               &  0.58$\pm$0.16    &  80$\pm$9     \\
               &  K             &               &  0.54$\pm$0.18    &  58$\pm$11     \\
               &   6.7          &  272$\pm$3    &  2.3$\pm$0.8      &  50$\pm$10     \\       
	       &  12.0          &  399$\pm$4    &  2.6$\pm$0.9      &  52$\pm$11     \\       
               &  14.3          &  794$\pm$7    &  3.1$\pm$0.9      &  62$\pm$9      \\
               &                &               &                   &                \\  
IRAS15250+3609 &   6.7          &  84$\pm$2     &  4.6$\pm$1.8      &  153$\pm$10      \\       
               &  12.0          & 109$\pm$3     &  6.3$\pm$1.5      &  151$\pm$7     \\
               &  14.3          & 200$\pm$5     &  4.3$\pm$1.4      &  164$\pm$9     \\
               &                &               &                   &                \\  
Mrk~273         &  J             &               &  0.0$\pm$0.35     &                \\
               &  K             &               &  0.42$\pm$0.30    &                \\
               &  6.7           & 153$\pm$2     &  6.2$\pm$3.0      &  159$\pm$14     \\
               &  12.0          & 210$\pm$3     &  6.7$\pm$1.3      &  174$\pm$6     \\
               &  14.3          & 361$\pm$5     &  6.5$\pm$1.0      &  171$\pm$4     \\          
\hline
	\end{tabular}  
 	\end{center}

\end{table}

\section{Discussion}

For all ULIRGs we find significant polarisation in the range of $\sim
3 - 8 \%$. The highest polarisation is recorded in Mrk~231 which has a
clear AGN signature from other studies (e.g. Genzel et al. 1998).  The
lowest polarisation is found for Arp~220, which is generally thought to
be powered by star formation and where only very recently evidence for
a hidden quasar was found (Haas et al. 2001).

In Mrk~231 we can rule out a synchrotron interpretation as the
extrapolated non-thermal emission at $1$\,mm (Downes \& Solomon 1998)
is contributing less than 1\% of the 12\,$\mu$m emission. Also dust
scattering should not be an important mid IR polarisation mechanism
since the grains are tiny compared to the wavelength and their
scattering cross sections are very small (Smith et
al. 1995). Therefore the mid IR polarisation of Mrk~231 is most likely
due to dichroism of magnetically aligned spheroidal dust
grains. Carilli et al. (1998) presented high resolution radio images
of Mrk~231 which indicate the existence of a sub--kiloparsec disc with
its major axis oriented east-west and viewed at an orientation of
about $45^{\rm o}$. On an even smaller scale mapped with VLBI, there
is evidence for a compact core and a radio lobe, about 20\,pc from the
core, with the axis of the core--lobe system being approximately
perpendicular to the major axis of the sub-kiloparsec disc. The position angle of
the polarisation is approximately parallel to the major axis of the sub-kiloparsec disc.  If we
assume that the rotation axis of the putative dusty torus in Mrk~231 and the
sub--kiloparsec disc are the same and the magnetic field lies in the
plane of the disc (as in other disc like systems like the Milky Way),
then we expect the position angle of polarisation produced by dichroic
absorption to be parallel to this magnetic field direction as
observed. We cannot decide on the basis of these observations alone
whether the polarisation is emissive or absorptive although the fact
that the polarisation angle is parallel to the plane of the disc is
tempting us to favor the latter mechanism. With spectro--polarimetry
across the silicate feature one should be able to discriminate between
the two dichroic polarisation mechanisms.  Jones \& Klebe (1989) argue
that the wavelength dependence of the optical to near IR polarisation
of Mrk~231 is not consistent with a screen of aligned grains nor with
synchrotron radiation. They favor scattering with dilution as the
prime polarisation mechanism.  The source of dilution in the optical
could be starlight whereas in the near IR may be due to a power--law
component.  We note that scattering of UV to near IR radiation by dust
flowing outward along the North--South VLBI axis would give rise to
polarisation with a position angle of roughly $90^{\rm o}$. This could
explain the relative agreement between the optical and mid IR position
angles despite the fact that the polarisation mechanism is different.

For Arp~220, Jones \& Klebe (1989) argue that the near IR polarisation
is due to extinction through aligned grains. The main argument is that
the wavelength dependence of polarisation is consistent with normal
interstellar polarisation as known from the Milky Way.  The position
angle of polarisation in Arp~220 is aligned with the position angle of
the $5'' \times 1.6''$ CO disc observed by Downes \& Solomon
(1998). The polarisation angle follows also the dust lane seen in the
optical. If the magnetic field structure is parallel to the disc, such
an orientation of the polarisation angle is expected if we interpret
the detected 6--15 $\mu$m polarisation as due to dichroic absorption
by aligned dust grains in the disc. ISO observations and models
(Siebenmorgen et al. 1999) of Arp~220 result in high dust extinction
estimates of $A_{\rm V} \sim 54$\,mag to the center of the disc. This
may be converted to a 14$\mu$m optical depth of $\tau_{14 \mu \rm m}
\sim 2.5$ using extinction profiles (e.g. Fig.~12 in Kr\"ugel \&
Siebenmorgen 1994). With perfectly aligned spinning spheroids of
silicate or amorphous carbon, and for such high optical depths, the
degree of mid IR polarisation is expected to be high (Siebenmorgen \&
Kr\"ugel 2000). On the contrary, the observed polarisation is low, a
fact which points to inefficient grain alignment expected for
extremely disordered magnetic field structures. A similar
interpretation is given by Jones \& Klebe (1989) for the low near IR
polarisation.

IRAS~15250+3609 is classified as a LINER by Veilleux et al. (1995). Sopp
\& Alexander find a North--South extension in their 15\,GHz map, i.e
with roughly the same position angle as our polarisation
measurement. This again may give some indication to favor dichroic
absorption as the dominant mid IR polarisation mechanism.

Mrk~273 has a double nucleus with at least one of its nuclei known to
contain a type II AGN (e.g. Sargent 1972). In the mid IR the source
has been resolved into two components which contribute about equal
amounts to the total flux (Soifer et al. 2000). The northern component
shows a deeper silicate absorption feature than the southern source
and is also associated with a CO source which Downes \& Solomon (1998)
resolved into a compact source and a disc oriented East-West. The
position angle of this disc is perpendicular to the position angle of
the mid IR polarisation. This may give some indication that the mid IR
polarisation is due to dichroic emission.  Also we note that at a
distance of $\sim 150$\,Mpc we probe only the innermost 2\,kpc region
of the nuclei and mid IR polarimetry at higher resolution is certainly
required to clarify the situation.

The IR polarisation due to aligned dust grains shows, except for the
silicate band, a relative flat dependence with wavelength. So that
galaxies which show dichroic mid IR polarisation should also be
polarised in the far IR and submillimeter. Then according to the
dichroic model the far-IR emission should be polarised with a position
angle perpendicular to that seen in absorption.  This effect has been
observed by Hildebrand et al. (1984) in Orion. However, the far
IR/submm emission may well have significant contribution from star
burst activity or from the host galaxy so that the mid IR is probably
the best range to probe the AGN.

\begin{acknowledgements}

CIA is a joint development by the ESA Astrophysics Division and the
ISOCAM Consortium. The ISOCAM Consortium is led by the ISOCAM PI,
C. Cesarsky. 
\end{acknowledgements}

\end{document}